\documentclass[11pt,nofootinbib,preprintnumbers,amsmath,amssymb,aps,prc,onecolumn, showkeys,floatfix,superscriptaddress]{revtex4-1}
\usepackage{amsmath,graphicx,color}
\usepackage{hyperref}

\begin{document}

\title{Smallest drop of QGP: Thermodynamic properties of p-Pb collisions}

\author{Fernando G. Gardim}
\email{fernando.gardim@unifal-mg.edu.br}
\affiliation{Instituto de Ci\^encia e Tecnologia, Universidade Federal de Alfenas, 37715-400 Po\c cos de Caldas, MG, Brazil}

\author{Renata Krupczak}
\email{rkrupczak@physik.uni-bielefeld.de}
\affiliation{Departamento de Física, Centro de Ciências Físicas e Matemáticas, Universidade Federal de Santa Catarina, Campus Universitário Reitor João David Ferreira Lima, Florianópolis 88040-900, Brazil}

\author{Tiago Nunes da Silva}
\email{t.j.nunes@ufsc.br}
\affiliation{Departamento de Física, Centro de Ciências Físicas e Matemáticas, Universidade Federal de Santa Catarina, Campus Universitário Reitor João David Ferreira Lima, Florianópolis 88040-900, Brazil}

\date{\today}

\begin{abstract}
The extreme temperature and density conditions that arise in ultrarelativistic collisions of heavy nuclei enable the formation of the most fundamental fluid in the universe, the deconfined phase of quantum chromodynamics known as quark-gluon plasma. Despite the extensive experimental evidence gathered over the last decade for the production of quark-gluon plasma in colliding systems such as Au-Au and Pb-Pb, the formation of quark-gluon plasma in the collision of smaller systems such as p-Pb remains an open question. In this study, we describe the evolution of matter formed in p-Pb collisions at 5.02 TeV using a state-of-the-art hybrid model based on viscous relativistic hydrodynamics. We study the thermodynamic properties of the medium and the final state observables. Our results are compared with experimental data and first-principles lattice quantum chromodynamics calculations. The results support the formation of a collective phase of strongly interacting matter in high-multiplicity p-Pb collisions.  
\end{abstract}

\maketitle

\section{Introduction}

The strong interaction of nature is described by quantum chromodynamics (QCD), the theory of color-charged particles: quarks and gluons. This theory predicts that under extreme conditions, matter undergoes a transition from a phase in which quarks and gluons are confined in hadrons to a deconfined phase in which they are free to propagate. This extreme phase of strongly interacting matter is called the quark-gluon plasma (QGP) ~\cite{Cabibbo:1975exp, Collins1975super}, and can be produced by the collision of nuclei at ultrarelativistic velocities in large particle accelerators ~\cite{BRAHMS:2004adc, PHOBOS:2004zne, PHENIX:2004vcz, STAR:2005gfr,Muller:2012zq,Foka:2016vta}.

The aftermath of a collisional event is described by the so-called standard model of heavy-ion collisions~\cite{Heinz:2013th}. Immediately after the collision, a lumpy and out-of-equilibrium state of matter is initially created. After a short time, it reaches near thermal equilibrium and becomes describable by relativistic viscous hydrodynamics ~\cite{Israel:1976tn,Israel:1979wp, Denicol:2012cn,Molnar:2013lta}. As the fluid QGP expands, its density and temperature decrease, and when it finally reaches the QCD confinement transition, quarks and gluons are hadronized. The resulting hadrons move toward the detectors of the experiment, experiencing both elastic and inelastic interactions. Numerical modeling of these various processes reproduces each of the above stages in a component of a hybrid model \cite{Petersen:2008dd}.

The main experimental evidence for the formation of QGP in a liquid phase is the collective behavior observed in large colliding systems such as those of Au-Au and Pb-Pb nuclei both at RHIC \cite{ BRAHMS:2004adc, PHOBOS:2004zne, PHENIX:2004vcz, STAR:2005gfr} and at the LHC \cite{Muller:2012zq,Foka:2016vta}. However, analyses in small systems such as proton-proton (p-p), p-Pb and d-Au, which were used as control measurements, showed similar behaviour as in large nuclei, suggesting that a hydrodynamic description of strongly interacting matter formed in these systems may also be possible~\cite{Weller2017one}. For example, measurements of the elliptic flux $v_2\{2\}$ in p-p~\cite{CMS:2016fnw} and p-Pb~\cite{ ATLAS:2013jmi} showed that $v_2\{2\}$ is non-zero. Later measurements of the elliptic flow fluctuations agreed with those expected for a liquid \cite{Ollitrault:2009ie} only for p-Pb and Pb-Pb collisions, while this was not the case for p-p~\cite{CMS:2016fnw}.

Another feature of a relativistic fluid is the correlation between its initial shape and the anisotropic flow in the final state~\cite{Teaney:2010vd, Gardim:2014tya}. It has been suggested that by manipulating the initial geometry of small systems, such as p/d/$^3$He-Au, one can probe QGP formation by analyzing the flow patterns in these systems ~\cite{Nagle:2013lja}. Relativistic viscous hydrodynamic predictions are consistent with experimental results of the PHENIX collaboration~\cite{PHENIX:2018lia} for these systems.


With the exception of jet quenching~\cite{CMS:2016xef}, all evidence supporting the formation of QGP in large-system collisions has been observed in p-Pb and d-Au collisions as well (for a review, see~\cite{Nagle:2018nvi}). Nevertheless, a physical mechanism can explain this phenomenon. The smaller medium produced results in a shortened average path. This is due to the short lifetimes of the medium, which prevents energy loss for partons with high $p_T$ as they travel through it.  At the same time, hydrodynamization for short-lived systems remains under discussion, even if hydrodynamics can be applied to small systems \cite{Romatschke:2016hle,Soloviev:2021lhs}. In the course of this debate, hydrodynamic simulations have been carried out for p-Pb systems \cite{Romatschke:2015gxa, Bozek:2014era} that only partially agree with the experimental data, even when a global Bayesian analysis is used to extract the maximum a posteriori model parameters~\cite{Nijs:2020roc, Moreland:2018gsh}.

The validity of the hydrodynamic picture for describing collisions of small systems remains an open question, which can be summarised in the daring question: "What is the smallest possible size of the most fundamental fluid in nature?". In this work, we investigate this question using a new approach to ultrarelativistic heavy ion collisions: We extract the thermodynamic properties of the hot, strongly interacting matter formed in p-Pb collisions simulated with a state-of-the-art hybrid model, combine them with experimental data, and test their consistency with first-principle calculations from lattice QCD.

%
\section{Thermodynamics from relativistic hydrodynamics}
\label{sec:thermo}
%
Thermodynamic properties of matter produced in ultrarelativistic heavy ion collisions, such as its effective temperature and entropy, can be obtained from the final observables if the matter is described by relativistic hydrodynamics. Remarkably, it was shown in~\cite{Gardim:2019xjs} that the thermodynamic properties obtained from the experimental data agree with the first-principles lattice QCD~\cite{Borsanyi:2013bia} calculations.

\begin{figure}
\begin{center}
\includegraphics[width=0.5\textwidth]{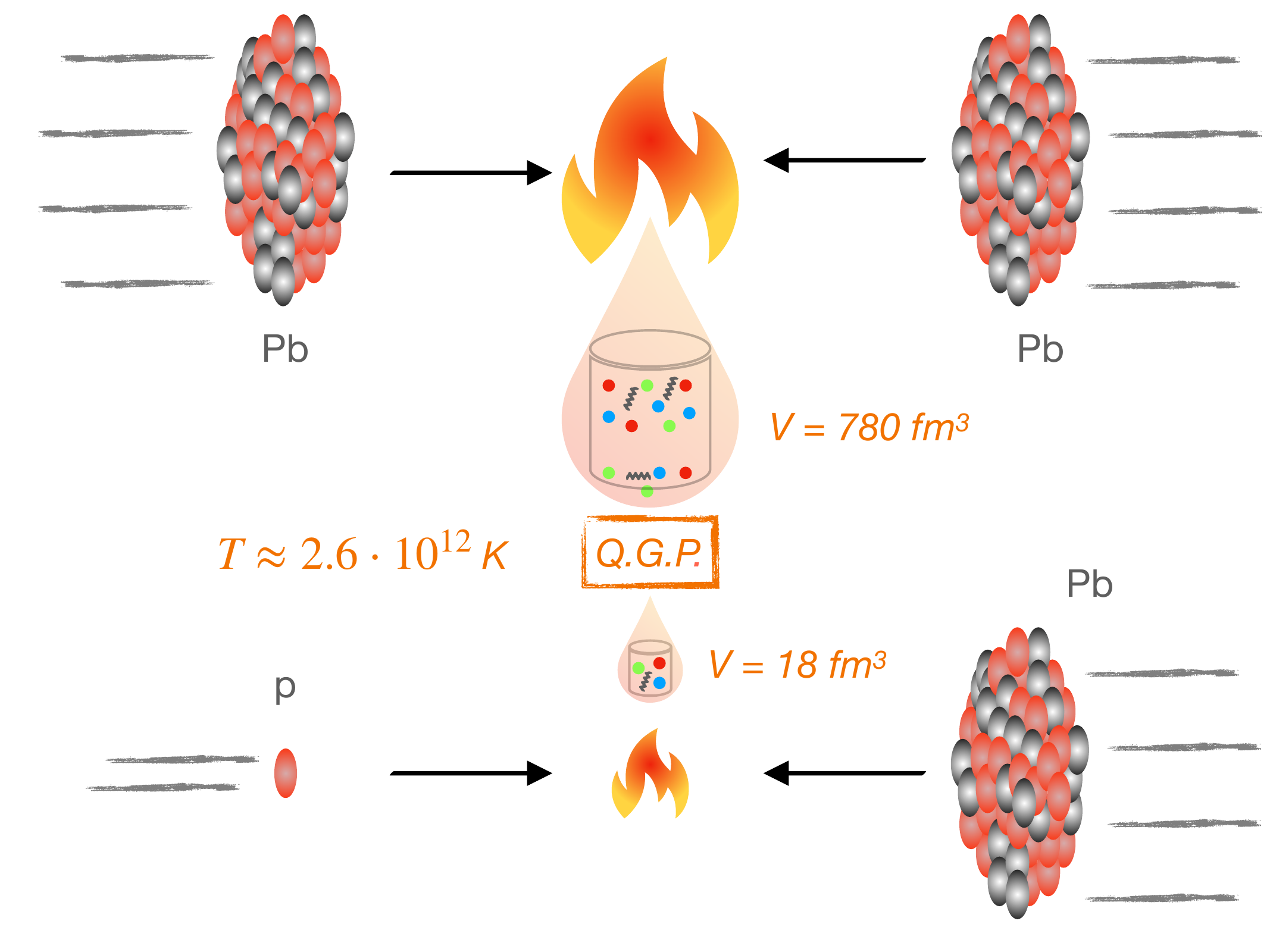}
\end{center}
\caption{Cartoon of nuclear collisions at ultrarelativistic energies: In nuclear collisions with very high energies, the matter produced can be seen as a fluid at rest, with dimensions billions of times smaller than a drop of water and temperatures of about trillions of Kelvin. What is new is that this also applies to systems even smaller than Pb-Pb, as in the collision of protons with lead.}
\label{fig:cartoon}
\end{figure}

The main idea developed in \cite{Gardim:2019xjs} is the following: The strongly interacting matter produced in heavy-ion collisions expands hydrodynamically in vacuum and cools when diluted. Its temperature is not uniform throughout the QGP and evolves according to the space-time history of the fluid. However, from the total energy $E$ and total entropy $S$ of the fluid at freeze-out, it is possible to capture the entire space-time evolution at an effective temperature $T_{\rm eff}$ and an effective volume $V_{\rm eff}$ (see the illustration in Figure \ref{fig:cartoon} ), analogous to a uniform fluid at rest having the same $E$ and $S$.

%
\begin{eqnarray}
  \label{eq:effective}
S=\int_{\rm f.o.} d\sigma_\mu s u^{\mu} &=&s(T_{\rm eff}) V_{\rm eff}, \cr
E=\int_{\rm f.o.} d\sigma_\mu T^{0\mu} &=&\epsilon(T_{\rm eff}) V_{\rm eff},
\end{eqnarray}
where $T^{\mu\nu}$ is the stress-energy tensor of the fluid, $u^{\mu}$ is the fluid four-velocity, the integrals are computed over the freeze-out hypersurface~\cite{Ollitrault:2008zz}, and $s$ and $\epsilon$ are the entropy and energy density, respectively. Taking the ratio $E/S$, one eliminates $V_{\rm eff}$ and can solve the resulting equation for $T_{\rm eff}$, with the aid of the equation of state.

According to hydrodynamic simulations of Pb-Pb nuclei collisions at center-of-mass energies $\sqrt{s_{NN}} = 5.02 \text{ TeV}$, it was found that the mean transverse momentum of charged particles $\langle p_t\rangle$ and $T_{\rm eff}$ are related by $\langle p_t\rangle\approx 3 T_{\rm eff}$ ~\cite{Gardim:2019xjs, Gardim:2020sma}. Moreover, this relationship has been shown to be independent of the transport coefficients and system centrality. This result is consistent with expectations, since for an ideal massless gas at very high energies the transverse momentum coincides with the energy per particle, i.e. $\epsilon=3T$. Combining these results with the experimentally measured value of $\langle p_t \rangle$, one obtains the effective temperature of $T_{\rm eff} = 222 \pm 9$ MeV, which confirms the production of the QGP in heavy-ion collisions.

To access the equation of state of the fluid, another thermodynamic quantity is required. It is known that for a relativistic fluid the entropy density is directly proportional to the particle density ~\cite{Ollitrault:2008zz, Gardim:2019xjs}, even if the hydrodynamic expansion is viscous ~\cite{Giacalone:2020ymy}. It is thus possible to determine the effective entropy density of a system by considering into account
\begin{eqnarray}
  \label{eq:s_effec}
s(T_{\rm eff}) = \frac{1}{V_{\rm eff}}\frac{\mathcal{S}}{N_{\rm ch}}\frac{dN_{\rm ch}}{dy},
\end{eqnarray}
where the multiplicity per unit rapidity $dN_{\rm ch}/dy$ is measured experimentally and the entropy per particle $\mathcal{S}/N_{\rm ch}$ can be obtained from experimental data~\cite{Hanus:2019fnc}. The quantity $s/T^3$ taken from Pb-Pb data is also in agreement with lattice QCD results \cite{Bazavov:2018mes}.

\section{Hydrodynamic description of \MakeLowercase{p}-P\MakeLowercase{b} COLLISIONS}
\label{sec:hydroppb}
%
To investigate which is the smallest drop QGP that can behave hydrodynamically, we adapt the strategy described above to the study of small colliding systems. Namely, we study $p$-Pb collisions at the center-of-mass energy $\sqrt{s_{NN}} = 5.02 \text{ TeV}$, since this is the smallest system for which experimental results suggest the formation of QGP. 

The thermodynamics of ultrarelativistic collisions is extracted for the first time from event-by-event minimum-bias hybrid simulations. This is the state-of-the-art simulation of heavy ion collisions.  As an extension of the framework proposed in \cite{Gardim:2019xjs}, we simulate the evolution of the matter formed in the collisions with a hybrid model that also accounts for the pre-equilibrium dynamics and hadronic interactions after the thermal freeze-out. This model provides a better description of the experimental data and is now the state-of-the-art simulation technique for heavy-ion collisions.

Two-dimensional initial conditions at time $\tau=0^+$ were generated using the parametric model $\mathrm{T}_{\mathrm{R}} \mathrm{ENTo}$ ~\cite{moreland2015alt}. These were free-streamed~\cite{broniowski2009free, liu2015pre} up to a hydrodynamic time $\tau_{f.s.}$ to model the pre-equilibrium dynamics of the system. The resulting initial hydrodynamic conditions were evolved using relativistic viscous hydrodynamics with MUSIC ~\cite{schenke2010hyd} assuming boost invariance. The hydrodynamic evolution terminates at a freeze-out temperature $T_\text{switch}$. The freeze-out hypersurface is stored for the calculation of the effective thermodynamic quantities. It is also sampled using the Cooper-Frye framework~\cite{Cooper1974sin}, including corrections for shear and bulk viscosity. The sampled hadrons are then evolved using the UrQMD~\cite{Bass1998mic} hadronic cascade model, taking into account the possible scatterings and decays. The final particles are used to calculate observables such as the final multiplicity and the mean transverse momentum, which are used in the analysis.

The set of hydrodynamic equations is closed using the HotQCD equation  of state for the QGP \cite{Bazavov:2018mes}, matched \cite{eosmaker} at low temperatures to a hadronic resonance gas (HRG). The particle content of the HRG is the same as for UrQMD. All parameters used in the simulation were the maximum a posteriori values obtained in the global Bayesian analysis presented in~\cite{Moreland:2018gsh}. 

\section{Results}
\label{sec:results}
%
\subsection{Effective quantities and final states observables from hybrid modeling}

\begin{figure}
\begin{center}
\includegraphics[width=0.7\textwidth]{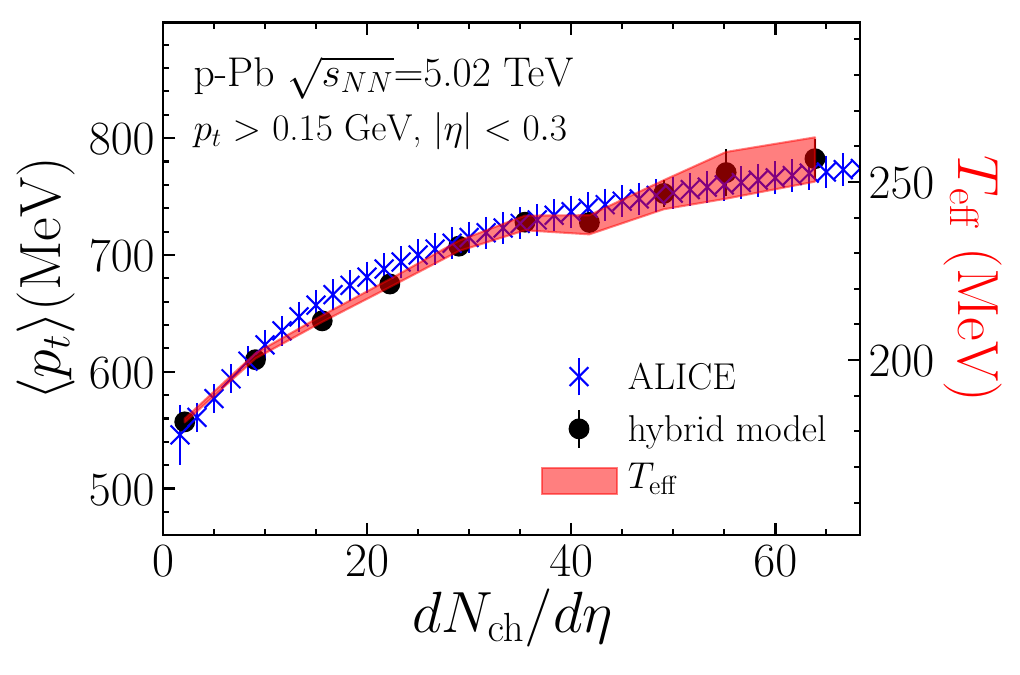}
\end{center}
\caption{Mean transverse momentum $\langle p_t\rangle$ of final charged particles produced in p-Pb collisions at $\sqrt{s}=5.02$ TeV from hybrid hydrodynamic simulations (black circle), and experimental measurements from the ALICE Collaboration (blue crosses) \cite{ALICE:2013rdo}. In both cases, $p_T> 0.15$ GeV and $|\eta|<0.3$. Effective temperature results are presented in the red band. These data can be used to establish a relation between transversal momentum and effective temperature.}
\label{fig:pttn}
\end{figure}

The mean transverse momentum in the final state calculated from the simulated events is shown in Figure~\ref{fig:pttn} as a function of the number of charged particles. The results are compared with experimental data from the ALICE collaboration~\cite{ ALICE:2013rdo}. In the same figure, we also plot the effective temperature calculated from Eq.~\ref{eq:effective} over the simulated freeze-out hypersurfaces. We have considered the same kinematic cuts used in the experimental analysis, namely $p_t > 0.15$ GeV and $|\eta|< 0.3$. The results show good agreement with the experimental data. Regardless of the final-state multiplicity, we obtain the relation
\begin{eqnarray}
  \label{eq:pt}
\langle p_t\rangle= 3.05\,T_{\rm eff}.
\end{eqnarray}
This result is in remarkable agreement with Pb-Pb collisions~\cite{Gardim:2019xjs, Gardim:2020sma}, despite the different system sizes and numerical chain components. Ignoring the kinematic cuts and considering all charged particles, the proportionality coefficient in Eq.(\ref{eq:pt}) changes to 2.97.

\begin{figure}
\begin{center}
\includegraphics[width=0.7\linewidth]{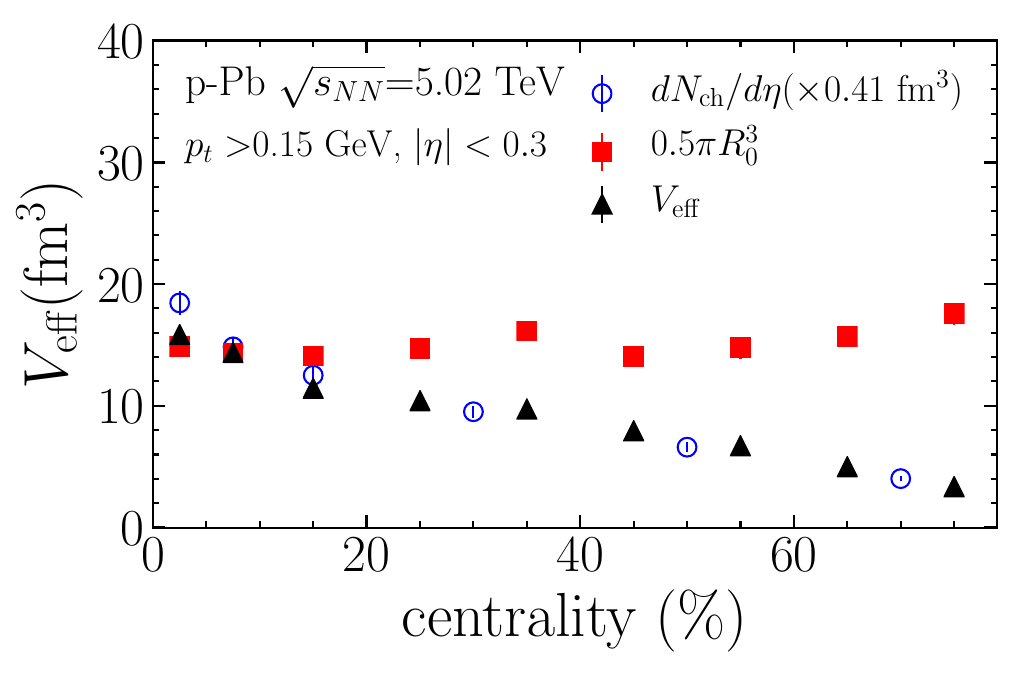} 
\end{center}
\caption{Effective volume as a function of centrality. The black triangles represent the effective volume calculated from the Eq. \ref{eq:effective}, the blue circles are the experimental data of multiplicity \cite{ALICE:2013rdo}  and the red squares represent the geometric idea for the effective volume based on the initial size, $R_0$. The effective volume is decreasing in the same proportion as the multiplicity, but the geometric volume is basically independent of the centrality.}
\label{fig:veff}
\end{figure}

Knowing the effective temperature, we can also compute the effective volume as a function of the number of charged particles, which is shown in Figure~\ref{fig:veff}. Similar to the Pb-Pb collisions, the effective volume decreases together with the final-state multiplicity, for more peripheral events.

The initial state size can be inferred from its radius, defined as 
\begin{equation}
    R_0 \equiv \sqrt{2\langle r^2 \rangle}
\end{equation}
where the above average is weighted by the entropy of the initial state. As can be seen in Figure~\ref{fig:veff}, the size of the initial state is roughly independent of the centrality for p-Pb systems, unlike what has been observed for Pb-Pb collisions. Since the initial size of the system remains the same for p-Pb events, regardless of how many particles are produced, the initial entropy density is smaller for peripheral events. This implies a decrease in the $\langle p_t\rangle$, which has been observed experimentally~\cite{ ALICE:2013rdo}. For Pb-Pb collisions, on the other hand, $\langle p_t\rangle$ is almost constant from central to peripheral collisions~\cite{ ALICE:2013rdo}. This is understandable since both the system size and the total entropy decrease when fewer particles are produced, i.e. the entropy density is roughly constant.

\subsection{Extracting thermodynamics from experimental data}
%
The results described in the previous subsection show that, assuming the validity of the hydrodynamic picture in ultra-relativistic collisions of small systems, it is possible to obtain the thermodynamic properties of this system from the observations of the final state. We will now use this fact to extract further thermodynamic quantities of the strongly interacting matter produced in p-Pb collisions at $\sqrt{s_{NN}}=5.02$ TeV.

First, we use Eq.~\ref{eq:pt} together with the $\langle p_t \rangle$ results from the ALICE Collaboration to obtain the effective temperature of the experimentally generated system. We use a factor of 3.05 due to the kinematic cuts used in the experiment. The resulting temperature range reached by the system for different multiplicities is
$$
179 < T_{\rm eff} < 257 \text{ MeV}.
$$
We note that this result differs from the Pb-Pb case, for which the effective temperature is approximately independent of the multiplicity and has the approximate value $T_{\rm eff}\approx 222$ MeV.

Another interesting result emerges when $V_{\rm eff}$ and $T_{\rm eff}$ are combined. Although the effective volume of both p-Pb and Pb-Pb systems decreases with event multiplicity, only in the case of p-Pb is there a decrease in the system temperature, suggesting that we may not be accessing the same state at different sizes.

We now use Eq.~\ref{eq:s_effec} to obtain the entropy density of the p-Pb system in this temperature range and compare it with first-principles lattice QCD calculations. We can calculate $s/T^3$ as:

\begin{eqnarray}
  \label{eq:s_over_t}
\frac{s(T_{\rm eff})}{T^3_{\rm eff}} = \frac{1.19}{V_{\rm eff}}\frac{\mathcal{S}}{N_{\rm ch}}\frac{dN_{\rm ch}}{d\eta}\frac{3.05^3}{\langle p_T \rangle^3},
\label{eq:eosoverT}
\end{eqnarray}
where we use $(dN_{\rm ch}/dy)/(dN_{\rm ch}/d\eta)\simeq 1.19$ \cite{Hanus:2019fnc}, since $dN_{\rm ch}/dy$ is not a direct observable. The multiplicity results $dN_{ch}/d\eta$ are taken from the ALICE Collaboration \cite{ALICE:2013rdo}, and the effective volume $V_{\rm eff}$ is taken from our hybrid modeling results shown in Fig.~\ref{fig:veff}. The entropy per particle $\mathcal{S}/N_{\rm ch}$ was obtained from \cite{Hanus:2019fnc}, where the $3D$ $HBT$ radii improved results for large and small systems are similar, and we use $\mathcal{S}/N_{\rm ch} = 6.3 \pm 0.8$. We adopt a conservative scenario for estimating the uncertainties. On \cite{Gardim:2019xjs}, the influence of the transport coefficients and the different models of initial conditions is estimated to be $19\%$. Combining this with the error for $\mathcal{S}/N_{\rm ch}$ gives an error of $23\%$ for our estimate of $s(T_{\rm eff})/T^3_{\rm eff}$.


Figure 4 shows the entropy density obtained from experimental data using Eq \ref{eq:eosoverT}, and compares it with the results from \textit{ab initio} calculations in QCD. The essential point in this comparison is that $T_{\rm eff}$ and $s(T_{\rm eff})$ are related through the equation of state of the fluid by construction, see Eq. \ref{eq:effective}, i.e., assuming the matter created in p-Pb collisions can be described by the standard model of relativistic nuclear collisions. The thermodynamics obtained from experimental data are in agreement with the Lattice QCD results, when $N_{\rm ch} \ge 11$ for the kinematic cuts used, corresponding to the $30\%$ of events with the highest multiplicity. These results support the formation of the QGP in collisions of small systems and suggest that it has the same collective properties observed in collisions of large systems.

\begin{figure}
\begin{center}
\includegraphics[width=0.7\linewidth]{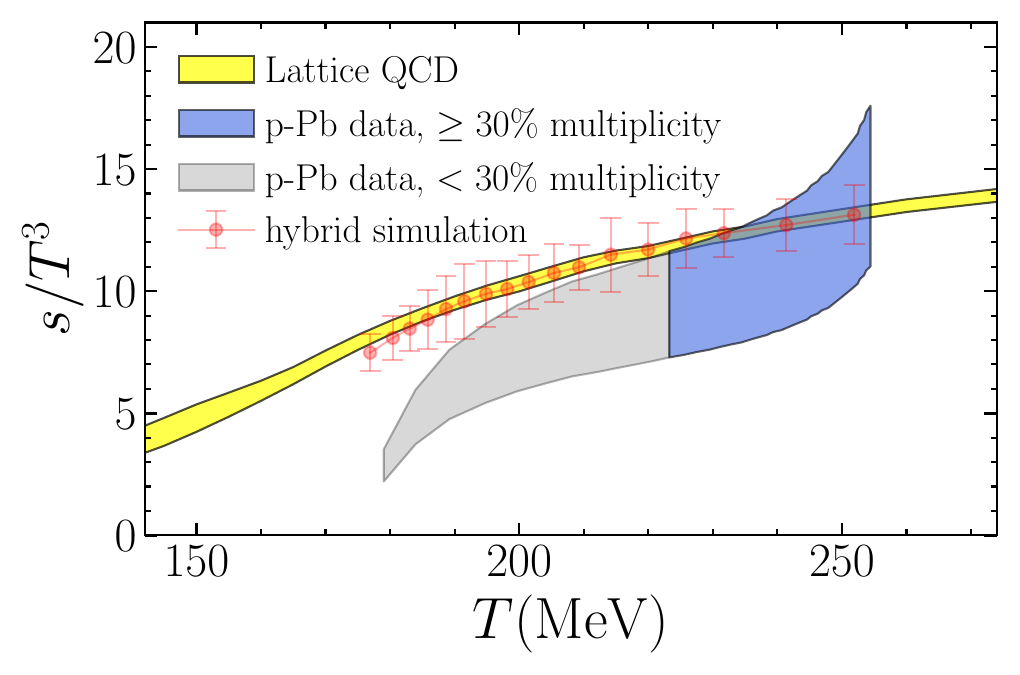}
\end{center}
\caption{Entropy density scaled by $T^3$ as a function of temperature. The yellow band represents the Lattice QCD calculation~\cite{Borsanyi:2013bia}, where the width of the band is determined by its uncertainty. The blue and gray band are the results of the calculation of Eq. \ref{eq:s_over_t}, and the width of the band is the uncertainty from the experimental data used in the calculation. The blue band agrees with the lattice QCD calculations, note the agreement occurs for the $30\%$ of events with the highest multiplicity. As a consistency check, we also show the curve  $s(T_\mathrm{eff})/T_\mathrm{eff}^3$ obtained when using, in Eq. (5), the results for $dN_{ch}/d\eta$ obtained from the hybrid simulations. }
\label{fig:eos1}
\end{figure}   

It is also possible to estimate the number of active degrees of freedom within the system by modeling the QGP as a massless ideal gas with Boltzmann statistics. The particle density is then given by $n=\nu T^3/\pi^2$, where $\nu$ is the number of degrees of freedom (color, flavor, spin). If the number of hadrons produced is equal to the number of quarks and gluons and one considers that only two-thirds of the hadrons are charged, the particle density in the effective volume is $n=1.5(dN_{\rm ch}/d\eta)/V_{\rm eff}$. For the collisions with high-multiplicity, in agreement with the lattice QCD results, $T_{\rm eff} > 220$ GeV, $dN_{\rm ch}/d\eta > 20$~\cite{ ALICE:2015nuz} and $V_{\rm eff} > 10.9$~fm$^3$, leading to $n > 2.7$~fm$^{-3}$ and $\nu > 19$. This large number shows that the color degrees of freedom are active, or, in other words, that a deconfined state should be produced.

\section{Conclusion}
\label{sec:conclusion}
%
This was the first time that state-of-the-art hybrid modeling of ultra-relativistic collisions was used to obtain thermodynamic quantities of strongly interacting matter. It is possible to extract the effective temperature and entropy density of the system from the number of charged particles in the final state and the mean transverse momentum measured in large ultra-relativistic colliders for p-Pb collisions. The results show that the thermodynamic quantities agree with the lattice QCD calculations only for the $30\%$ of p-Pb events with the highest multiplicities. Events with smaller multiplicities do not show the properties expected for QGP. The temperature that delimits the region where p-Pb behaves like a QGP droplet is approximately the same as previously found for Pb-Pb collisions, namely $T_{\rm eff}\approx 220$ MeV. However, for p-Pb events, the volume of the system is 45 times smaller than for central Pb-Pb events.

One may argue that the differences between the hybrid model and the experimental data for p-Pb might be able to be fixed by making the chain parameters more accurate. Given the robustness of our findings about the correlation between $\langle p_t \rangle$ and $T_{\rm eff}$, our conclusion regarding the validity of the hydrodynamic framework for these systems is more robust than a simple comparison of usual final observables to experimental data.

Another quantity of thermodynamic interest is the speed of sound $c_s$, which can be obtained by momentum variations with respect to multiplicity, $c_s^2=d\ln \langle p_t \rangle/d\ln(dN/d\eta)$, with fixed centrality, i.e. same effective volume \cite{Gardim:2019xjs}. In this calculation, the proportionality coefficients between observables and thermodynamics, which are the largest sources of errors, cancel each other out. Extracting this quantity would depend on measurements of the average transverse momentum of p-Pb systems at different energies. Recently, the ALICE collaboration has presented preliminary results for p-Pb on the dependence of the transverse momentum as a function of the number of charged particles for two different energies, 5.02 TeV and 8.16 TeV \cite{ALICE:2022xip}, however, it is not possible to obtain a $c_s$ at this point as no dependence on centrality was shown.

These results should be taken as an important caveat when extracting QGP transport coefficients from Bayesian studies: if small systems are included in the study, only a centrality range consistent with QGP formation should be considered in the analysis. 

Although there is already experimental data for p-p indicating the absence of collective behavior, we can qualitatively infer from this study that in the case of fluid behavior for p-p, the value of the effective temperature would be higher than in the case of p-Pb. This would imply a decrease in the ratio $s/T^3$ and bring it under lattice QCD, confirming the experimental results. However, a rigorous comparison with the standard model of relativistic collisions is still needed before we can rule it out completely. This is deferred for future study.

\section*{Acknowledgments}
FGG and TNdS are supported by CNPq (Conselho Nacional de Desenvolvimento Cientifico) through the INCT-FNA grant 312932/2018-9 and the Universal grant 409029/2021-1. FGG was supported by CNPq 306762/2021-8 and FAPESP grant 2018/24720-6. RK is supported by a CAPES (Coordenação de Aperfeiçoamento de Pessoal de Nível Superior) fellowship.

\end{document}